Jonas Himmel, Max Ehrhardt, Matthias Heinrich, Malte Röntgen, Alexander Szameit, and Tom A. W. Wolterink*

# Eigenmodes of latent-symmetric quantum photonic networks

**Abstract:** We investigate the impact of latent symmetries on the dynamics of photonic systems and their eigenmodes. Residing solely within the eigenspectral domain, latent symmetries are not visible in real space, yet promise intriguing new ways to engineer the functionality of photonic systems. We study the eigenmodes of a 9-site latent-symmetric photonic network and find that an anti-symmetric input state is fundamentally precluded from populating so-called singlet sites. Furthermore, arbitrary extensions of the system at these sites do not break its latent symmetry. Therefore anti-symmetric excitations cannot leave the initial system, which can be leveraged e.g. for the storage of information. This holds true for both single-photon states, or classical light, as well as both distinguishable and indistinguishable two-photon quantum states. Latent symmetries introduce a powerful new set of tools to the design of systems with desired functionality on any nanophotonic platform, paving the way for applications in photonic information processing.

## 1 Introduction

The design and implementation of photonic systems plays a key role in photonic information processing, both in the classical and quantum domain. While photonic devices may be realized on a wide variety of technological platforms ranging from nanophotonic metasurfaces to integrated photonics, most of these structures can be described in terms of coupled modes, which are then tailored to realize the desired functionality. In order to understand and predict the dynamics within such system, symmetries are of paramount importance. Spatial symmetries, such as permutation symmetry are well studied in terms of their influence on the system's dynamics, and help to design e.g. networks for perfect transfer of quantum states [1], [2], [3], [4]. Meanwhile, symmetries in $k$-space serve to establish and classify topological insulators and superconductors [5], and $PT$-symmetry can be applied for enhanced sensing [6], [7] or mode-selective laser cavities [8], [9] in non-Hermitian settings. While the presence of such conventional symmetries imposes corresponding symmetry constraints in the spectral domain, spectral symmetries do not necessarily show up in real space, yet still influence the behaviour of the studied system. This is demonstrated by the concept of cospectrality [10], which has been recently introduced in graph theory and gave rise to the concept of so-called latent or 'hidden' symmetries [11], [12]. Latent symmetries are not visible in real space, and exclusively manifest in the spectrum of the photonic system. They give rise to e.g. flat bands [13], topology [14] or non-Hermiticity [15], and enable functionalities such as the transfer of states [16], [17], [18].

* **Corresponding author: Tom A. W. Wolterink**, Institute for Physics, University of Rostock, Rostock, Germany; tom.wolterink@uni-rostock.de; https://orcid.org/0000-0002-3591-8044
**Jonas Himmel, Max Ehrhardt, Matthias Heinrich and Alexander Szameit:** Institute for Physics, University of Rostock, Rostock, Germany
**Malte Röntgen:** Eastern Institute for Advanced Study, Eastern Institute of Technology, Ningbo, Zhejiang 315200, People's Republic of China; https://orcid.org/0000-0001-7784-8104

In this work we focus on the impact of latent symmetries on the eigenmodes of a photonic system, and their effect on the dynamics of single and two-photon excitations within the system. Starting with single-photon dynamics, we demonstrate that anti-symmetric excitations of the latent symmetric network sites lead to destructive interference at so-called singlet sites [19]. Intriguingly, this holds true even when the system is arbitrarily extended at these singlet sites. In other words, an anti-symmetric input state can be 'stored' in the initial system without being able to leave it. Expanding this approach to the evolution of distinguishable and indistinguishable two-photon states, we show how the resulting dynamics are influenced by the latent symmetry.

## 2 Latent-symmetric network

To illustrate the effects of the latent symmetry on its eigenmodes and dynamics, we study the nine-site latent-symmetric system shown in Fig. 1a. A linear optical system containing $N$ coupled modes can be described by the Hermitian Hamiltonian

$$\mathbf{H} = \sum_i^N \beta_i |i\rangle\langle i| + \sum_{i,j}^N \kappa_{i,j} |i\rangle\langle j|,$$

with the real on-site potential $\beta_i$ and the couplings $\kappa_{i,j}$ between sites $i$ and $j$. The evolution of an arbitrary single-photon input state, or equivalently an excitation with classical light, is then given by

$$|\psi(t)\rangle = \mathbf{U}(t)|\psi(t=0)\rangle,$$

with the time-evolution operator $\mathbf{U}(t) = \exp(-it\mathbf{H})$. By decomposing the Hamiltonian in terms of its eigenvalues $\lambda_r$ and associated eigenmodes $|\lambda_r\rangle$ to $\mathbf{H} = \sum_r \lambda_r |\lambda_r\rangle\langle\lambda_r|$, the time-evolution operator can be written as

$$\mathbf{U}(t) = \sum_r e^{-it\lambda_r} |\lambda_r\rangle\langle\lambda_r|.$$

Being fully determined by the eigenmodes and eigenvalues, the dynamics of the system are therefore subject to the influences of symmetries occurring in the spectral regime. One example of such a symmetry is the *cospectrality* of two sites of the system. Two sites of a network are considered cospectral if the two networks obtained by removing either one of them share the same set of eigenvalues [20]. While this is obviously the case whenever the two sites are permutation symmetric, this condition can also be fulfilled if they are not. Indeed, two sites can be cospectral without obeying any corresponding symmetry in the spatial domain [11]. Networks with at least one pair of such sites are referred to as *latent symmetric*. In the example shown in Figure 1a, sites $u = 2$ and $v = 6$ exhibit a latent symmetry.

Recently, the term *strong cospectrality* has been introduced [10]. If $u$ and $v$ are cospectral and $\mathbf{H}$ only has simple (non-degenerate) eigenvalues, $u, v$ are strongly cospectral and fulfil

$$\langle\lambda_r|u\rangle = \pm\langle\lambda_r|v\rangle,$$

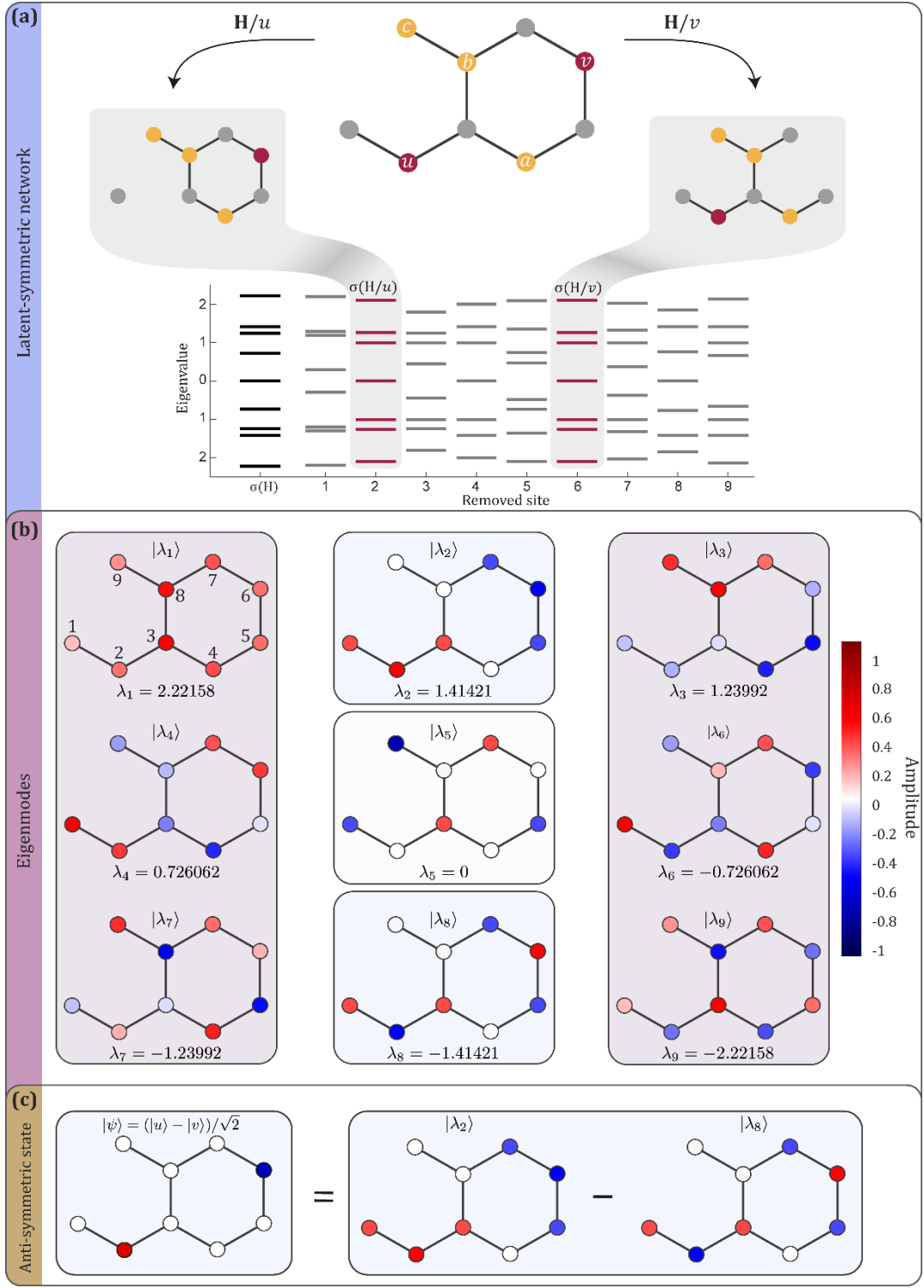

**Fig. 1: (a)** A nine-site system with two latent symmetric sites $u, v$ and singlet sites $a, b, c$. While there is no permutation symmetry between $u$ and $v$, their respective removal yields two graphs $H/u$ and $H/v$ with the same eigenvalue spectrum $\sigma(H/u) = (H/v)$, and hence the two network sites are latent symmetric. **(b)** The nine eigenmodes and eigenvalues of the system. The colormap represents the amplitude of the eigenmodes on the respective network site. Sites $u = 2$ and $v = 6$ have definite parity in all modes, hence they are strongly cospectral. Even (odd) parity eigenmodes have red (blue) backgrounds, whereas the eigenmode that vanishes on $u$ and $v$ has white background. **(c)** An anti-symmetric input state on $u$ and $v$ only populates the two odd parity eigenmodes, which have vanishing amplitude on the singlet sites $a, b, c$

with the canonical basis vectors $|u\rangle, |v\rangle$ that have a value of one at positions $u$ and $v$, while remaining strictly zero everywhere else. This means that for every eigenmode the components representing the sites $u$ and $v$ are of same amplitude with either same or opposite sign. In other words, all eigenmodes have either even or odd parity on these sites.

The eigenmodes of the studied network are displayed in Figure 1b. Six of them have even parity on $u$ and $v$ (red background), two have odd parity (blue background) and one eigenmode has vanishing amplitude on the two sites (white background). This parity implies that $u$ and $v$ are not only cospectral, but also strongly cospectral. The system's ground mode $|\lambda_1\rangle$ has the same sign of amplitude on all sites. The second mode $|\lambda_2\rangle$ is the first excited mode where the sign of one side of the system switched. It is followed by the second and third excited mode and so on. Since the system is represented by a bipartite graph, the eigenvalues are symmetric around zero and come in $\pm$ pairs [21], while the eigenmodes of opposite eigenvalue have same but alternating amplitudes. Additionally, if a site $w$ has the same "distance" to the two cospectral sites $u$ and $v$, it is referred to as a *singlet site* [19].

Intriguingly, such sites possess the property of allowing for arbitrary extension of the network without breaking the latent symmetry between $u$ and $v$. It is worth noting that each singlet site may have either even or odd parity with respect to the cospectral sites, further information on this topic is found in [19]. Our system (Figure 1a) has three singlet sites $a = 4$, $b = 8$ and $c = 9$ (highlighted yellow), all of even parity with respect to $u$ and $v$.

## 3 Single-photon dynamics

As a signature of latent symmetry, the behaviour of anti-symmetric excitations at the two latent-symmetric network sites is determined by two important properties. Firstly, Theorem 4 of [19] states that the amplitude of odd eigenmodes always vanishes on even singlet sites. Secondly, an (anti) symmetric input state in $u$ and $v$, $|\psi\rangle = (|u\rangle \pm |v\rangle)/\sqrt{2}$, only excites the (odd) even parity eigenmodes of the system. A combination of these two statements leads to the conclusion that an anti-symmetric input state in $u$ and $v$ will never populate any singlet sites. Considering the odd eigenmodes of the studied system in Figure 1b, it is easily seen that they indeed all have vanishing amplitude on the singlet sites $a$, $b$ and $c$. Figure 1c shows that an anti-symmetric input state only excites the odd parity eigenmodes of the system, $|\lambda_2\rangle$ and $|\lambda_8\rangle$, and thus never populates the singlet sites $a$, $b$ and $c$. We emphasize that this holds true for all systems with a pair of latent-symmetric sites and at least one singlet site, which makes it a signature of latent symmetry.

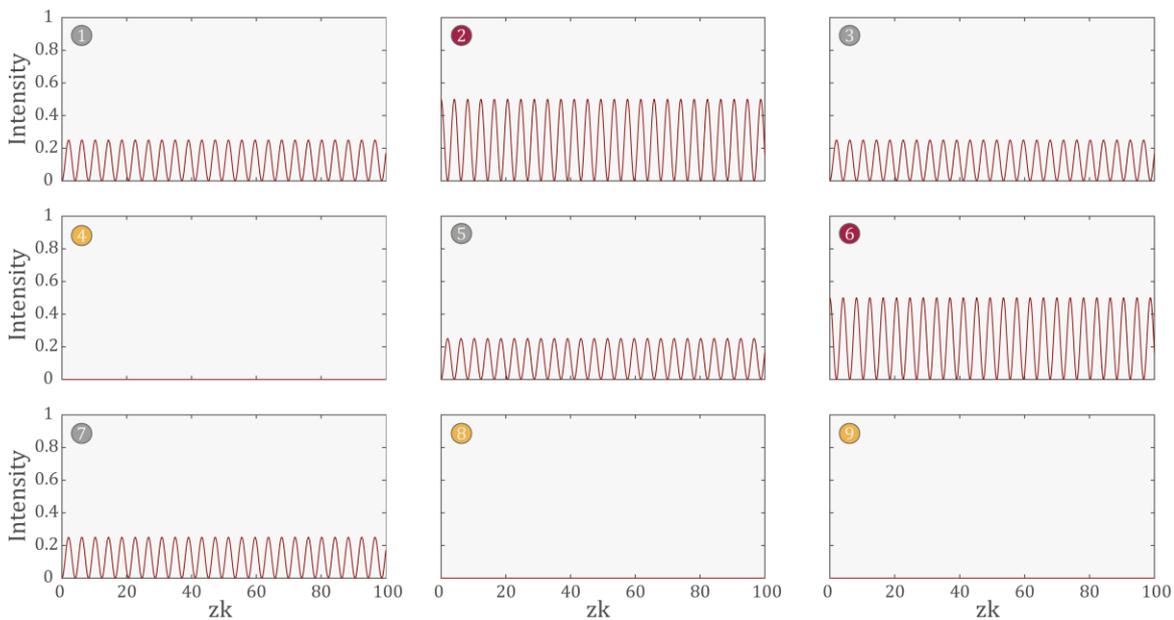

**Fig. 2:** Evolution of a classical excitation through the latent symmetric network. Shown is the intensity as function of the effective coupling distance defined as the product of propagation distance $z$ and coupling constant $k$.

In an integrated photonic implementation of the system, which maps its time evolution to the propagation direction of the light, this property can be observed using a classical anti-symmetric excitation on $u$ and $v$. The classical intensity distribution of the system then corresponds to the average photon number $n$ for single-photon excitations. The singlet sites $a$, $b$ and $c$ will then always stay dark for all propagation distances (Figure 2). Since only two eigenmodes are excited, the time-dependent amplitude at any of the six other networks sites follows a cosine behaviour.

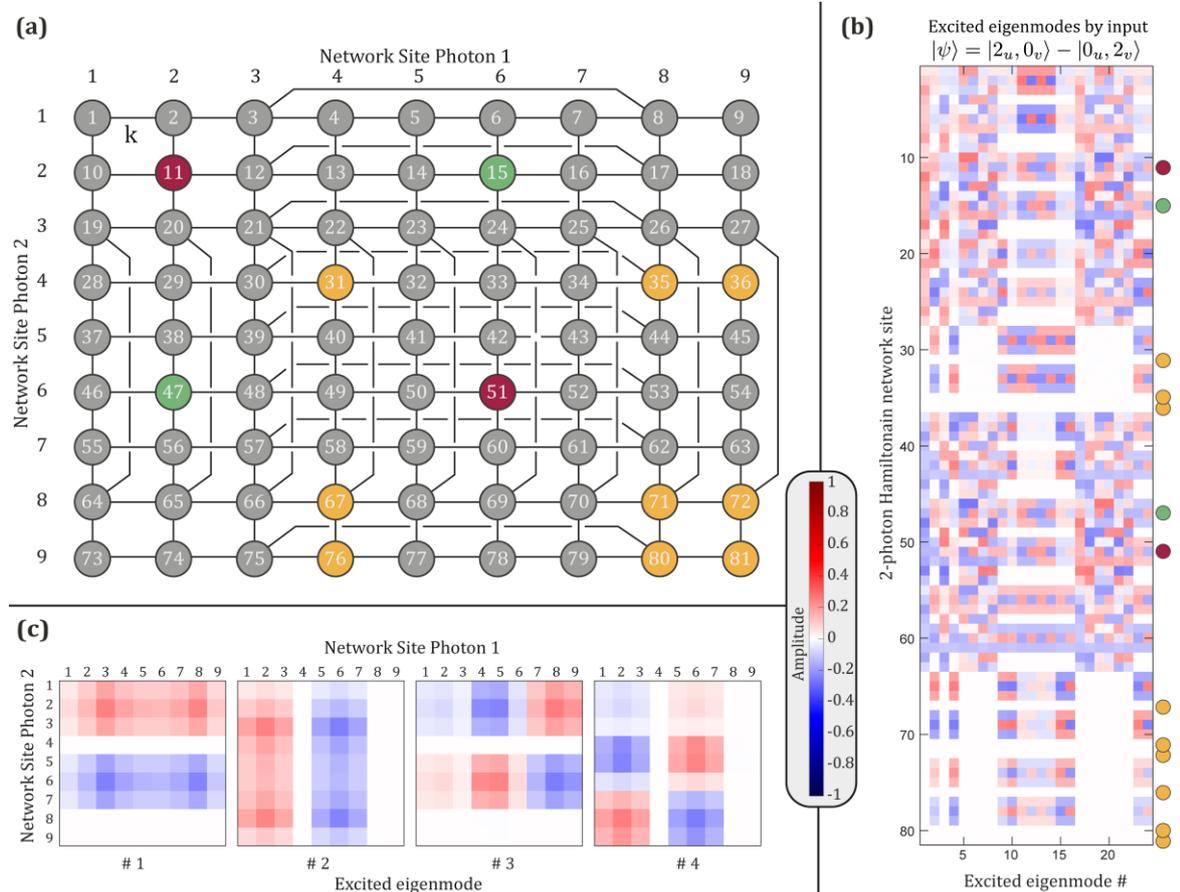

**Fig. 3: (a)** Two-photon Hamiltonian $H_{Dist}$ of the studied system can be represented by a $9 \times 9 = 81$ site graph. Each of its sites corresponds to the position of the two photons in the original graph at a time. There are now two latent symmetric pairs and nine singlet sites. **(b)** An anti-symmetric two-photon input state on site 11 and 51, $|\psi\rangle = (|2_u 0_v\rangle - |0_u 2_v\rangle)/\sqrt{2}$, excites the 24 odd parity eigenmodes out of the total 81 eigenmodes. The amplitude on the singlet sites is zero for all of them, while both latent-symmetric pairs have odd parity amplitudes. An anti-symmetric input state on site 15 and 47 would excite the same modes. **(c)** The first four of the 24 eigenmodes excited by an anti-symmetric input reshuffled to the shape of the $9 \times 9$ site graph.

## 4 Two-photon dynamics

To explore the dynamics of a multi-particle state, we apply the above two statements to the evolution of a distinguishable two-photon state through the studied system. The dynamics of a such a state can be described by the two-photon Hamiltonian $\mathbf{H}_{\text{Dist}} = \mathbf{H} \oplus \mathbf{H}$, where $\oplus$ denotes the Kronecker sum. $\mathbf{H}_{\text{Dist}}$ corresponds to a network with $N^2 = 81$ sites, as depicted in Figure 3a. Each site now represents a two-photon correlation event. Site 32 e.g. corresponds to photon one residing at site 5 while photon two is located at site 4. The evolution of a two-photon input state is then given by the unitary operator $\mathbf{U}_{\text{Dist}} = e^{-it\mathbf{H}_{\text{Dist}}}$. Again, it is determined by the interference of its eigenmodes and thus their symmetries. If $\Lambda$ is the diagonal matrix containing the single-photon eigenvalues $\lambda_r$, then the two-photon eigenvalues are obtained via the diagonal entries of $\Lambda_{\text{Dist}} = \Lambda \oplus \Lambda$. Similarly, one obtains the eigenmodes of the two-photon Hamiltonian via the matrix $V_{\text{Dist}} = V \otimes V$, if $V$ is the matrix with the single-photon eigenmodes $|\lambda_r\rangle$ as columns, with $\otimes$ the Kronecker product.

The resulting Hamiltonian $\mathbf{H}_{\text{Dist}}$ (Figure 3a) features two latent-symmetric pairs and nine singlet sites. The first latent symmetric pair (red) refers to the case of both photons residing at site 2 or both at site 6 of the single-photon graph $\mathbf{H}$. The second pair (green) refers to one photon being at site 2, the other one at site 6 of $\mathbf{H}$ and vice versa. The singlet sites of $\mathbf{H}_{\text{Dist}}$ refer to the cases of both photons being at (possibly different) singlet sites of $\mathbf{H}$, so to the correlation of both photons being at one of the singlet sites $a$, $b$ or $c$. The sites corresponding to one photon being at a singlet site and one photon at another site of $\mathbf{H}$ do not become singlet sites of $\mathbf{H}_{\text{Dist}}$.

Applying the same definitions as in the single-photon case, we study the evolution of an anti-symmetric input state, $|\psi_1\rangle = (|2_u 0_v\rangle - |0_u 2_v\rangle)/\sqrt{2}$ or $|\psi_2\rangle = (|1_u 1_v\rangle - |1_u 1_v\rangle)/\sqrt{2}$,

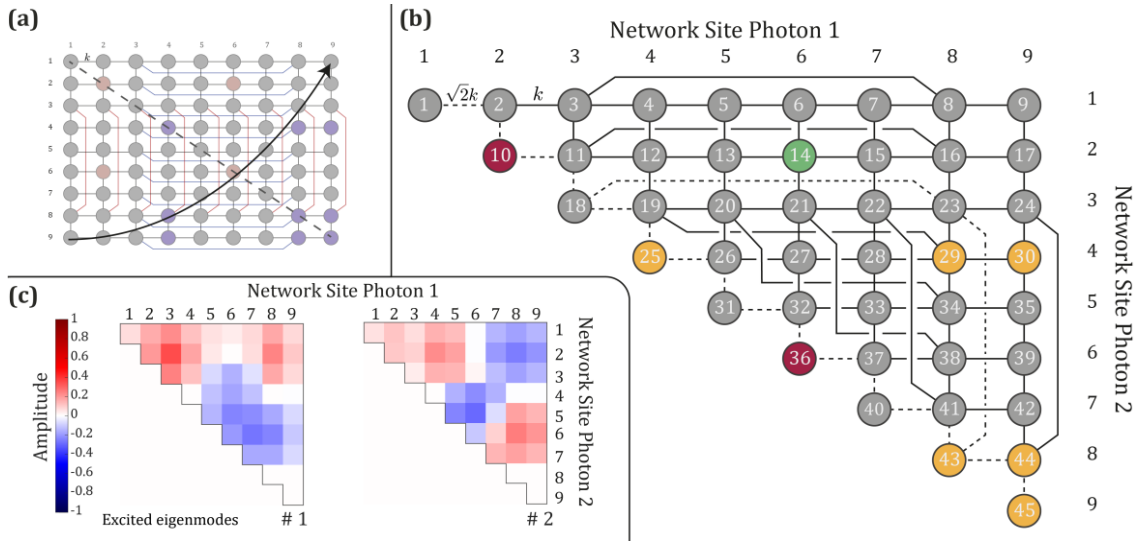

**Fig. 4:** (a) Folding the two-photon Hamiltonian for distinguishable photons $H_{Dist}$ across the diagonal yields the two-photon Hamiltonian or indistinguishable photons $H_{Indist}$. (b) The two latent symmetric sites (red) as well as the single sites (yellow) on the diagonal remain constant. The off-diagonal singlet sites merge from six to three without changing their symmetry properties, while the off diagonal latent symmetric sites (green) interfere destructively in the odd parity eigenmodes and becomes a singlet site. The dashed line indicates a coupling of $\sqrt{2}k$, the solid line a coupling of $k$. (c) The first two out of twelve eigenmodes excited by an anti-symmetric input state $|\psi\rangle = (|2_u 0_v\rangle - |0_u 2_v\rangle)/\sqrt{2}$ reshuffled to the shape of the Hamiltonian, which can be directly constructed by 'folding' the four eigenmodes shown in Figure 3c. As expected, singlet sites are zero, latent symmetric sites have odd parity.

which correspond to exciting either site 11 and 51 or 15 and 47 of $\mathbf{H}_{\text{Dist}}$. Both input states exclusively populate the odd parity eigenmodes of the two-photon Hamiltonian $\mathbf{H}_{\text{Dist}}$ as shown in Figure 3b for $|\psi_1\rangle$. These modes always have zero amplitude at the two-photon singlet sites. When exciting the latent-symmetric sites of $\mathbf{H}_{\text{Dist}}$ in an anti-symmetric fashion, no correlations will be observed between the singlet sites of $\mathbf{H}$: both photons will never be found at some combination of sites $a$, $b$ and $c$. However, correlations between singlet sites and non-singlet sites of $\mathbf{H}$ do indeed occur: one photon can populate a singlet site $a$, $b$ or $c$ while the other photon is at a non-singlet site. In Figure 3c the first four of the excited modes are shown, reshuffled to the shape of the Hamiltonian $\mathbf{H}_{\text{Dist}}$. The first mode is separable into photon 1 being in mode $|\lambda_1\rangle$ and photon 2 in mode $|\lambda_2\rangle$, as one would expect for distinguishable non-interacting particles. Furthermore, it has the same shape as the second one and the third as the fourth, just for the particles swapped.

As a next step, we consider indistinguishable photons and study how latent symmetry affects quantum interference of photons. In that case, the space of possible correlations decreases, since the outcome of the first photon being at site $m$ and the second photon being at site $n$ cannot be distinguished from the first photon being at site $n$ and the second photon being at site $m$. Intuitively, the indistinguishable two-photon Hamiltonian $\mathbf{H}_{\text{Indist}}$ is obtained [22], by diagonally 'folding' one half of the Hamiltonian onto the other (Figure 4a). Specifically, the pairs of rows and columns that will become indistinguishable are added, the rows and columns that connect to two photons in the same mode, for normalisation, are multiplied by $\sqrt{2}$ and the constructed matrix is then divided by two. We thus end up with the $((N \times N) + N)/2$-site Hamiltonian (Fig. 4b). An indistinguishable anti-symmetric input state $|\psi\rangle = (|2_u 0_v\rangle - |0_u 2_v\rangle)/\sqrt{2}$ again excites only the twelve odd parity eigenmodes of $\mathbf{H}_{\text{Indist}}$ which are zero at the singlet sites. Upon examining the first two excited eigenmodes, it can be seen that the first one can be formed by individually folding either of the first two eigenmodes of distinguishable photon Hamiltonian, $\mathbf{H}_{\text{Dist}}$, over the diagonal onto itself. Similarly, the second mode can be formed by folding eigenmodes three and four in Figure 3c. As a result, we obtain half of the odd modes in total. While the latent symmetric site on the diagonal (both photons being found either at site $u$ or $v$) remains unchanged, the parity in the eigenmodes of the second latent-symmetric pair (one photon being at site $u$ and the other at $v$) causes destructive interference when folding the modes. As consequence, the remaining site becomes a singlet site (green), with zero amplitude in the odd eigenmodes. The site, corresponding to correlations on sites $u$ and $v$, thus also stays zero for all times, as a result of the quantum interference of indistinguishable photons.

## 5 Conclusion

In conclusion, we studied the dynamics of coupled optical modes in a latent-symmetric system for single photons as well as for (in)distinguishable two-photon states. It was shown theoretically that anti-symmetric single-photon excitations at the latent-symmetric sites are systematically precluded from populating the singlet sites. Since the system can be arbitrarily extended at these sites without breaking the symmetry, this could be e.g. used for the storage of information – an anti-symmetric input state would never leave the initial latent-symmetric system. We furthermore demonstrated that also all singlet sites of the constructed distinguishable two-photon Hamiltonian stay zero for all times. By extending this theory to the evolution of an indistinguishable two-photon state, we found quantum interference as a result of the transformation of two latent symmetric network sites to one singlet site. This way we contributed to a deeper understanding of latent symmetries and their influence on quantum light, which can be applied to implement e.g. state transfer networks.

While we here used a network of waveguides to illustrate the concept of latent symmetry, the concept itself is rooted in graph theory and can therefore readily be applied to any coupled-mode system in photonics, such as nanophotonic metasurfaces and metamaterials, coupled resonators, or optomechanical systems. The eigenspectral domain offers an additional route for harnessing symmetries to control the behaviour of photonic systems, without relying on real space or $k$-space, which furthermore connects well to the new approaches in non-Hermitian [14] and topological physics [15], [24]. Thus, latent symmetries add a powerful tool to design photonic metasurfaces with desired functionality for applications in photonic information processing, sensing, and metrology [25], [26]. It will be exciting to explore the interplay of latent symmetries and multiphoton quantum states, as that directly increases the dimensionality of the underlying graph. An excellent question would be whether it is possible to induce or suppress latent symmetries by introducing more photons and their (in)distinguishability. That way, latent symmetries may support engineering quantum-photonic metasurfaces [27] that shape the quantum behaviour of light.

**Research funding**: AS acknowledges funding from the Deutsche Forschungsgemeinschaft (grants SZ 276/9-2, SZ 276/19-1, SZ 276/20-1, SZ 276/21-1, SZ 276/27-1, and GRK 2676/1-2023 'Imaging of Quantum Systems', project no. 437567992). AS also acknowledges funding from the Krupp von Bohlen and Halbach Foundation as well as from the FET Open Grant EPIQUS (grant no. 899368) within the framework of the European H2020 programme for Excellent Science. AS and MH acknowledge funding from the Deutsche Forschungsgemeinschaft via SFB 1477 'Light–Matter Interactions at Interfaces' (project no. 441234705). TAWW is supported by a European Commission Marie Skłodowska-Curie Actions Individual Fellowship (project no. 895254).